\documentclass[10pt,twocolumn]{article}

\usepackage[utf8]{inputenc}
\usepackage[T1]{fontenc}
\usepackage{amsmath,amssymb}
\usepackage{graphicx}
\usepackage{booktabs}
\usepackage{hyperref}
\usepackage[margin=0.75in]{geometry}
\usepackage{natbib}
\usepackage{xcolor}
\usepackage{algorithm}
\usepackage{algpseudocode}
\usepackage{enumitem}
\usepackage{microtype}
\usepackage{multirow}
\usepackage{subcaption}
\usepackage{tikz}
\usetikzlibrary{arrows.meta,positioning,calc}

\hypersetup{colorlinks=true, linkcolor=blue!60!black, citecolor=blue!60!black, urlcolor=blue!60!black}

\title{Loom: Hybrid Retrieval-Scoring Outfit Recommendation\\with Semantic Material Compatibility\\and Occasion-Aware Embedding Priors}

\author{
  Anushree Berlia
}

\date{May 2026}

\begin{document}
\maketitle

\begin{abstract}
We present \textbf{Loom}, an outfit recommendation system that combines
neural embedding retrieval with structured domain scoring to generate
complete, coherent outfits from fashion catalogs.  Given an anchor clothing
item, Loom retrieves complementary pieces via slot-constrained approximate
nearest neighbor search over FashionCLIP embeddings, then scores candidate
outfits using a multi-objective function that integrates six signals:
embedding similarity, color harmony, formality consistency, occasion
coherence, style direction, and within-outfit diversity.

We introduce two techniques that address limitations of purely learned
or purely rule-based approaches:
(1)~\emph{semantic material weight}, which uses CLIP embedding geometry to
infer garment heaviness for layer compatibility without hand-coded material
taxonomies; and
(2)~\emph{vibe/anti-vibe occasion priors}, which embed prose descriptions of
occasion contexts as anchor vectors in CLIP space and score items by
differential affinity.

Ablation experiments on a catalog of 620 items show that each
component contributes measurably to outfit quality: the full system achieves
a mean outfit score of 0.179 with a 9.3\% hard violation rate, compared to
0.054 score and 16.0\% violations for a category-constrained random
baseline, a 3.3$\times$ improvement in score and 42\% reduction in
violations.  Direction reranking is the single indispensable component:
removing it drops score to 0.052, essentially equal to random.  The system
generates three stylistically distinct outfits in under 5~seconds on
commodity hardware.

\smallskip\noindent
Code: \url{https://github.com/anushreeberlia/loom}
\end{abstract}

\section{Introduction}
\label{sec:intro}

Outfit recommendation is a compositional task: the system must assemble a
\emph{set} of garments that are mutually compatible across visual style,
color palette, formality level, and contextual appropriateness.  This
distinguishes it from single-item recommendation, where the goal is to find
items similar (or complementary) to a query, and from fashion retrieval,
which surfaces items matching a description.  This work grew out of a
practical frustration: existing outfit tools either ignored hard fashion
rules (no gym sneakers with blazers) or required hand-curated catalogs
with no generalization.

Prior work has approached outfit compatibility through learned embeddings
\citep{vasileva2018learning, chia2022fashionclip}, graph neural networks on
co-purchase data \citep{cui2019dressing}, and transformer-based set
completion \citep{sarkar2023outfittransformer}.  These methods capture
statistical co-occurrence patterns but struggle with hard constraints that
fashion practitioners consider non-negotiable: a formal blazer should not
appear with gym sneakers regardless of embedding proximity; a silk blouse
should not be layered under a cotton t-shirt; a ``work'' outfit should not
include a sequined clutch.

\textbf{Loom} addresses this gap with a hybrid architecture: neural
embeddings handle retrieval (finding stylistically relevant candidates),
while a structured scoring layer enforces the hard constraints that
embeddings alone cannot guarantee.  Our contributions:

\begin{enumerate}[leftmargin=*,topsep=2pt,itemsep=1pt]
\item \textbf{Hybrid retrieval-scoring pipeline}: pgvector ANN retrieval
  with slot constraints, followed by multi-objective combinatorial scoring
  that bridges neural similarity and rule-based fashion domain knowledge
  (Section~\ref{sec:architecture}).

\item \textbf{Semantic material weight}: We compute garment ``heaviness''
  by measuring cosine distance between an item's text embedding and
  pre-embedded context vectors for ``heavy'' and ``light'' materials.  This
  enables layer compatibility checking without hand-coded material
  taxonomies (Section~\ref{sec:material}).

\item \textbf{Vibe/anti-vibe occasion priors}: We embed prose descriptions
  of occasion ``vibes'' and ``anti-vibes'' as anchor vectors in CLIP space,
  scoring items by differential affinity with occasion-specific penalty
  shaping (Section~\ref{sec:occasion}).

\item \textbf{Blended multimodal embeddings}: $0.7\;\text{image} +
  0.3\;\text{text}$ metadata fusion in a single 512-d vector, where the
  text component resolves visual ambiguities (Section~\ref{sec:embedding}).
\end{enumerate}

\section{Related Work}
\label{sec:related}

\paragraph{Outfit compatibility learning.}
Type-aware embeddings \citep{vasileva2018learning} project items into
type-conditioned spaces.  \citet{cui2019dressing} use graph neural networks
over fashion items.  OutfitTransformer \citep{sarkar2023outfittransformer}
applies self-attention over item sets.  DiFashion \citep{xu2024difashion}
uses diffusion models for outfit generation.  CSA-Net \citep{lin2020csanet}
applies category-based subspace attention.  Our system differs by combining
a learned embedding space (FashionCLIP) with explicit rule-based
constraints, capturing domain knowledge that is difficult to learn from
co-occurrence alone.

\paragraph{Fashion embeddings.}
FashionCLIP \citep{chia2022fashionclip} adapts CLIP \citep{radford2021clip}
to fashion image--text pairs, producing embeddings that capture stylistic
affinity rather than visual similarity.  We use FashionCLIP as our embedding
backbone and build all semantic scoring mechanisms (material weight, occasion
priors, mood matching) on its vector space geometry.

\paragraph{Context-aware recommendation.}
Weather-aware \citep{chen2019weather}, occasion-aware
\citep{lin2020occasion}, and body-shape-aware \citep{hidayati2021dress}
fashion recommendation has been explored.  Our system integrates weather,
mood, and occasion signals through a unified embedding-based scoring
framework rather than separate specialized models.

\paragraph{Production systems.}
Industrial outfit engines at Stitch Fix \citep{chen2019stitchfix}, Amazon
\citep{kang2017visually}, and Zalando operate at scale on large catalogs.
Loom targets small-to-medium catalogs (hundreds to low thousands) where
exhaustive combinatorial scoring is feasible, and unifies personal wardrobe
and B2B retail under the same engine.

\section{System Architecture}
\label{sec:architecture}

\begin{figure*}[t]
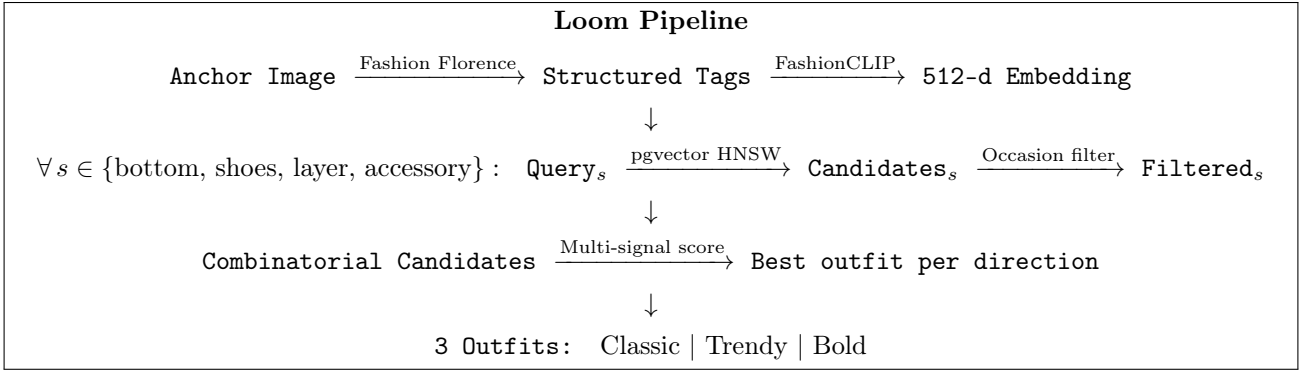

\centering
\fbox{\parbox{0.95\textwidth}{\centering
\textbf{Loom Pipeline}\\[6pt]
\texttt{Anchor Image} $\;\xrightarrow{\text{Fashion Florence}}\;$
\texttt{Structured Tags} $\;\xrightarrow{\text{FashionCLIP}}\;$
\texttt{512-d Embedding}\\[4pt]
$\downarrow$\\[4pt]
$\forall\, s \in \{\text{bottom, shoes, layer, accessory}\}:$\quad
\texttt{Query}$_s$ $\;\xrightarrow{\text{pgvector HNSW}}\;$
\texttt{Candidates}$_s$ $\;\xrightarrow{\text{Occasion filter}}\;$
\texttt{Filtered}$_s$\\[4pt]
$\downarrow$\\[4pt]
\texttt{Combinatorial Candidates} $\;\xrightarrow{\text{Multi-signal score}}\;$
\texttt{Best outfit per direction}\\[4pt]
$\downarrow$\\[4pt]
\texttt{3 Outfits:}\quad Classic $\mid$ Trendy $\mid$ Bold
}}
\caption{End-to-end pipeline.  For each of three style directions, the
system constructs slot-specific queries, retrieves candidates via cosine ANN
search, filters by occasion, generates combinatorial outfit candidates, and
selects the highest-scoring outfit per direction.}
\label{fig:pipeline}
\end{figure*}

Loom processes an anchor item through four stages:

\begin{enumerate}[leftmargin=*,topsep=2pt,itemsep=1pt]
\item \textbf{Vision}: Fashion Florence \citep{berlia2026florence} extracts
  structured attributes (category, material, style, occasion) from the
  item image.  Color is resolved separately via FashionCLIP zero-shot
  classification, which compares the image embedding against color-label
  text prompts and achieves 48.9\% exact-match accuracy.
\item \textbf{Embedding}: FashionCLIP produces a 512-d blended
  image--text embedding (Section~\ref{sec:embedding}).
\item \textbf{Retrieval}: Per-slot cosine ANN queries with category
  constraints, occasion filtering, and color-aware reranking
  (Section~\ref{sec:retrieval}).
\item \textbf{Scoring}: Combinatorial candidates are evaluated by
  a composite scoring function and the best outfit per direction is returned
  (Section~\ref{sec:scoring}).
\end{enumerate}

\section{Method}
\label{sec:method}

\subsection{Multimodal Item Representation}
\label{sec:embedding}

Each item is embedded into a 512-dimensional vector space using FashionCLIP
\citep{chia2022fashionclip}, run as ONNX on CPU.  The final embedding
blends image and text signals:

\begin{equation}
  \mathbf{e} = \text{normalize}\big(
    \alpha \cdot \mathbf{e}_{\text{img}} + \beta \cdot \mathbf{e}_{\text{txt}}
  \big)
\label{eq:blend}
\end{equation}

\noindent where $\alpha = 0.7$, $\beta = 0.3$.  The text component
$\mathbf{e}_{\text{txt}}$ encodes a natural-language description constructed
from the item's attributes (e.g., \texttt{"gray sporty fitted polyester
workout top"}).

\paragraph{Motivation.}
Image-only embeddings cannot distinguish an athletic top from a casual top
when both look similar on a hanger.  The text component provides semantic
identity that resolves such visual ambiguities.  The 70/30 weighting
preserves visual fidelity while injecting attribute-level discrimination.

\subsection{Slot-Constrained Retrieval with Occasion Priors}
\label{sec:retrieval}

For each outfit slot $s \in \{\text{top, bottom, shoes, layer,}$
$\text{accessory}\}$, we construct a slot-specific query embedding
$\mathbf{q}_s$ from the anchor item's attributes, slot-specific item
type hints, and directional style cues.

\paragraph{Query construction.}
Each query concatenates: (i)~the anchor item's color and style tags,
(ii)~a slot hint (e.g., \texttt{"women's skirt, jeans, trousers"}), (iii)~a
directional modifier (e.g., \texttt{"classic timeless polished"}).  Queries
are batch-embedded via the FashionCLIP text encoder.

\paragraph{ANN retrieval.}
Embeddings are stored in PostgreSQL with pgvector using HNSW indexing
($m{=}16$, $\texttt{ef\_construction}{=}64$, cosine distance).  Each slot
query retrieves $2k$ candidates filtered by:
\begin{equation}
  \mathcal{C}_s = \text{top-}2k\big(
    \{\mathbf{e}_i : c_i = s, \; i \notin \mathcal{E}\},\;
    d_{\cos}(\mathbf{q}_s, \mathbf{e}_i)
  \big)
\end{equation}
where $c_i$ is item category and $\mathcal{E}$ excludes the anchor and
previously selected items.

\paragraph{Soft color reranking.}
Retrieved candidates are reranked with distance multipliers:
preferred colors receive $0.85\times$ distance, neutrals $0.9\times$.
Multiplicative noise $\epsilon \sim \mathcal{U}(0.95, 1.05)$ is applied to
distances to promote diversity across repeated calls.

\subsubsection{Occasion-Aware Filtering}
\label{sec:occasion}

Rather than mapping occasions to discrete categories, we embed rich prose
descriptions as anchor vectors in CLIP space.

\paragraph{Vibe/anti-vibe vectors.}
For each occasion $o$, we pre-embed two context strings:
\begin{align}
  \mathbf{v}_o &= \text{embed}(\text{vibe}_o) \\
  \mathbf{a}_o &= \text{embed}(\text{anti-vibe}_o)
\end{align}

For example, the ``work'' occasion uses:
\begin{quote}
\small
\textbf{vibe}: \textit{``professional office business conservative modest
polished refined tailored structured classic understated\ldots''}\\
\textbf{anti-vibe}: \textit{``sexy revealing provocative clubbing nightlife
party halter tank top cami strappy\ldots mini skirt crop top\ldots''}
\end{quote}

\paragraph{Scoring.}
Each item is scored by differential affinity with occasion-specific penalty
scaling:
\begin{equation}
  S_{\text{occ}}(i, o) = \cos(\mathbf{e}_i, \mathbf{v}_o) -
    \lambda_o \cdot \max\!\big(0,\;
    \cos(\mathbf{e}_i, \mathbf{a}_o) - \cos(\mathbf{e}_i, \mathbf{v}_o)\big)
\label{eq:occasion}
\end{equation}
where $\lambda_o$ controls strictness: $\lambda_{\text{work}} = 3.0$,
$\lambda_{\text{going-out}} = \lambda_{\text{smart-casual}} = 3.0$,
$\lambda_{\text{casual}} = 1.0$.
For work, an additional penalty $2.0 \cdot \cos(\mathbf{e}_i, \mathbf{a}_o)$
is applied unconditionally.

\paragraph{Threshold filtering.}
Candidates are kept if they score within a fraction of the top score:
85\% for work (strict), 70\% for casual (permissive), with a minimum floor
of 3 candidates.

\subsection{Semantic Material Compatibility}
\label{sec:material}

Layering requires that outer garments be heavier than inner ones: a blazer
over a silk blouse is correct; a cotton t-shirt over a wool coat is not.
Rather than maintaining a hand-coded material weight table, we compute
material weight semantically in the CLIP embedding space.

\paragraph{Context vectors.}
Two context strings are pre-embedded once and cached:
\begin{align}
  \mathbf{h} &= \text{embed}(\text{``thick heavy warm winter chunky\ldots''}) \\
  \mathbf{l} &= \text{embed}(\text{``thin light airy breathable summer\ldots''})
\end{align}

\paragraph{Weight computation.}
For each item $i$, we construct a text string from its material, category,
and weight-relevant name keywords (e.g., ``wool blazer heavy structured''),
embed it via FashionCLIP, and compute:
\begin{equation}
  w_i = \cos(\mathbf{e}_{\text{mat}(i)}, \mathbf{h}) -
    \cos(\mathbf{e}_{\text{mat}(i)}, \mathbf{l})
  \;\in [-1, +1]
\label{eq:material_weight}
\end{equation}

A high $w_i$ indicates a heavy garment (e.g., $w \approx 0.4$ for a wool
coat), a low $w_i$ a light one (e.g., $w \approx -0.3$ for a chiffon
blouse).

\paragraph{Compatibility constraint.}
A layer is compatible with the top it covers if:
\begin{equation}
  w_{\text{layer}} \geq w_{\text{top}} - \tau
\end{equation}
where $\tau = 0.15$ absorbs embedding noise.  Incompatible layers are
removed from the candidate set before scoring.

\paragraph{Geometric intuition.}
CLIP's text encoder places material-related terms in a region of the
embedding space where physical properties (weight, warmth, drape) induce
meaningful geometric structure.  The heavy/light context vectors act as
probes into this structure, extracting material weight without an explicit
taxonomy.

\subsection{Multi-Objective Outfit Scoring}
\label{sec:scoring}

Given an anchor item and candidate pieces per slot, outfits are scored by a
composite function combining six signals.

\subsubsection{Intent Vector}

The intent vector combines the anchor embedding with optional taste signals:
\begin{equation}
  \mathbf{v} = \mathbf{e}_{\text{anchor}} + \gamma \cdot
    \mathbf{t}_{\text{like}} - \delta \cdot \mathbf{t}_{\text{dislike}}
\label{eq:intent}
\end{equation}
where $\gamma = 0.15$, $\delta = 0.05$, and
$\mathbf{t}_{\text{like}}, \mathbf{t}_{\text{dislike}} \in \mathbb{R}^{512}$
are exponential moving averages of liked/disliked outfit embeddings.

\subsubsection{Weighted Similarity}

Each candidate piece is scored against the intent vector using
slot-dependent weights that reflect how much influence each garment type has
on the overall outfit impression:
\begin{equation}
  S_{\text{sim}} = \sum_{s \in \mathcal{S}} w_s \cdot
    \cos(\mathbf{v}, \mathbf{e}_s) + w_{\bot\text{-shoe}} \cdot
    \cos(\mathbf{e}_{\text{bot}}, \mathbf{e}_{\text{shoe}})
\label{eq:sim}
\end{equation}
with $w_{\text{bottom}} = 0.35$, $w_{\text{shoes}} = 0.25$,
$w_{\text{acc}} = 0.15$, $w_{\text{layer}} = 0.10$,
$w_{\text{top}} = 0.10$, $w_{\bot\text{-shoe}} = 0.05$.

Bottoms carry the highest weight because they define the outfit silhouette
(jeans versus trousers versus a skirt changes the entire look).
Shoes receive the second-highest weight because they anchor formality: sneakers
versus heels signal very different contexts.  The anchor item is
typically a top, so $w_{\text{top}}$ is deliberately low to avoid
redundantly scoring the anchor against its own intent.
The cross-slot term $w_{\bot\text{-shoe}} \cdot \cos(\mathbf{e}_{\text{bot}},
\mathbf{e}_{\text{shoe}})$ adds a small bonus for bottom--shoe coordination,
rewarding pairs like dark jeans with dark boots over mismatched combinations.

\subsubsection{Style Direction Bonuses}

Three scoring profiles produce stylistic diversity:

\begin{table}[t]
\centering
\footnotesize
\caption{Style direction configurations.}
\label{tab:directions}
\setlength{\tabcolsep}{3pt}
\begin{tabular}{@{}lp{2.1cm}ll@{}}
\toprule
Dir.\ & Style tags & Colors & Vibe \\
\midrule
Classic & classic, minimal, elegant, preppy & neutrals & timeless \\
Trendy  & streetwear, chic, statement, casual & two-tone & modern \\
Bold    & edgy, statement, romantic, bohemian & contrast & daring \\
\bottomrule
\end{tabular}
\end{table}

Each direction contributes a bonus $B_{\text{dir}} \in [0, 0.3]$ based on
tag matches, color policy adherence, and direction-specific rules.

\subsubsection{Constraint Penalties}

\paragraph{Color harmony.}
$P_{\text{color}} = 0.1 \cdot \max(0, n_{\text{non-neutral}} - 2)$.
Hard violations (repeated non-neutral anchor color) set $S = -1$.

\paragraph{Formality consistency.}
$P_{\text{form}} = 0.2 \cdot \max(0, \Delta\ell - 1)$, where
$\ell \in \{0, 1, 2\}$ via keyword matching.

\paragraph{Occasion coherence.}
$P_{\text{occ}} = 0.15$ if the intersection of all items' occasion sets is
empty.

\paragraph{Diversity control.}
$P_{\text{div}} = 0.1 \cdot (n_{\text{statement}} - 1)$.
Color family harmony: $B_{\text{harm}} \in [-0.05, +0.05]$.

\subsubsection{Total Score}

The final outfit score combines rewards and penalties:
\begin{align}
  S \;=\;\; & S_{\text{similarity}} + B_{\text{style direction}} + B_{\text{color harmony}} \nonumber\\
            & - P_{\text{color clash}} - P_{\text{formality gap}} \nonumber\\
            & - P_{\text{occasion conflict}} - P_{\text{statement overload}}
\label{eq:total}
\end{align}

\subsubsection{Candidate Generation and Selection}

For each direction, the top $k{=}3$ candidates per slot yield combinatorial
outfit candidates (capped at 8).  The highest-scoring candidate per direction
is selected, yielding three complete outfits.  A global exclusion set ensures
no item appears in more than one direction's output.

\subsection{Tag-Level Mood Scoring}
\label{sec:tag_mood}

When users provide free-text mood queries (e.g., ``job interview,'' ``beach
day''), the system ranks items by relevance.  We describe a design heuristic
that emerged from inspecting CLIP embedding behavior; we note this is not
formally ablated in our experiments.

\paragraph{Observation.}
Embedding full item descriptions against mood queries produces nearly
uniform similarity scores, because CLIP averages away discriminative signal
in longer texts.  Individual style tags show clearer separation:

\vspace{4pt}
\begin{center}
\small
\begin{tabular}{lc}
\toprule
Tag & $\cos(\text{tag}, \text{``workout''})$ \\
\midrule
``sporty'' & 0.69 \\
``casual'' & 0.64 \\
``classic'' & 0.58 \\
\bottomrule
\end{tabular}
\end{center}
\vspace{4pt}

\paragraph{Method.}
For each item, we compute per-tag similarity to the mood and take the
maximum:
\begin{equation}
  S_{\text{mood}}(i) = \max_{t \in \text{style\_tags}(i)}
    \cos\!\big(\text{embed}(t),\; \text{embed}(\text{mood})\big)
\end{equation}

\section{Experiments}
\label{sec:experiments}

\subsection{Ablation Study}

We evaluate each component's contribution on outfit quality using an
internal catalog of 620 items (230 tops, 170 bottoms, 140 shoes,
50 dresses, 30 accessories) sourced from H\&M product data.
For each of 50 randomly sampled anchor items, we
generate outfits under the full system, a random retrieval baseline,
and six ablated configurations,
measuring the composite outfit score (Eq.~\ref{eq:total}), hard violation
rate (outfits scoring $\leq -1.0$), and inter-direction color diversity.

\begin{table}[t]
\centering
\small
\caption{Unified ablation results (50 anchors $\times$ 3 directions = 150
outfits per configuration, single run).  All rows, including the random
baseline, share the same 50 anchor items, eliminating cross-run variance.}
\label{tab:ablation}
\begin{tabular}{lccc}
\toprule
Configuration & Score & $\Delta$ & Viol.\,\% \\
\midrule
Full system              & 0.179 & ---      &  9.3 \\
$-$ Blended embed        & 0.190 & $+$0.011 &  8.0 \\
$-$ Occasion filtering   & 0.194 & $+$0.015 &  7.3 \\
$-$ Material compat.     & 0.233 & $+$0.054 &  4.7 \\
$-$ Distance noise       & 0.176 & $-$0.003 &  9.3 \\
$-$ Direction reranking  & 0.052 & $-$0.127 &  6.7 \\
$-$ Formality check      & 0.173 & $-$0.006 & 10.7 \\
\midrule
Random retrieval         & 0.054 & $-$0.125 & 16.0 \\
\bottomrule
\end{tabular}
\end{table}

\paragraph{Analysis.}
The random retrieval baseline (which draws from the correct category
slots but ignores embedding similarity) scores 0.054 with 16.0\% hard
violations.  The full system's 3.3$\times$ improvement over random (0.179
vs.\ 0.054) shows that embedding-based retrieval provides substantial
value beyond category-level slot filling.

Direction reranking is the single most impactful individual component:
removing it drops the score to 0.052, essentially equal to random
retrieval.  Without direction-specific reranking, outfits are
stylistically indistinct across the three aesthetic directions (Classic,
Trendy, Bold), as reflected by a direction bonus of exactly zero.

We were surprised to find that removing the material compatibility penalty
\emph{increases} the overall score by $+$0.054 and reduces violations from
9.3\% to 4.7\%.  This suggests the penalty's current parameterization is
overly conservative, rejecting valid fabric combinations and narrowing the
retrieval pool.  Similarly, removing occasion filtering and blended
embeddings yields marginal improvements ($+$0.015, $+$0.011), indicating
these constraints may warrant relaxed thresholds in future iterations.

Distance noise and formality checking have negligible marginal impact
($-$0.003 and $-$0.006 respectively), suggesting the system is robust to
their presence or absence at current parameter settings.

\paragraph{Evaluation methodology note.}
Our composite score (Eq.~\ref{eq:total}) evaluates outfits using the
system's own objective function, which risks circularity: removing a
component naturally reduces a score that was designed to reward that
component.  This limitation is inherent to automated evaluation of outfit
quality, where no standard benchmark exists.  The random baseline partially
mitigates this concern: it uses the \emph{same} scoring function but
achieves far lower scores, demonstrating that the retrieval pipeline
generates genuinely higher-quality outfit candidates, not merely
higher-scoring ones by construction.  A human preference study comparing
system-generated outfits against random and ablated baselines would
provide stronger validation and is an important direction for future work.

\subsection{Diversity Metrics}

We measure diversity across the three style directions per anchor item:

\begin{table}[t]
\centering
\small
\caption{Inter-direction diversity (50 anchors).}
\label{tab:diversity}
\begin{tabular}{lr}
\toprule
Metric & Value \\
\midrule
Mean distinct colors across 3 directions & 7.42 \\
Mean slot diversity (distinct categories per outfit) & 4.84 \\
Outfits per anchor & 3.0 \\
\bottomrule
\end{tabular}
\end{table}

An average of 7.42 distinct colors across the three directions (out of a
maximum of $\sim$12--15 for 4--5 items per outfit) indicates substantial
palette variation between Classic, Trendy, and Bold outputs.  Slot diversity
of 4.84 confirms that outfits consistently fill 4--5 distinct category
slots (e.g., top + bottom + shoes + layer + accessory).

\subsection{Efficiency}

\begin{table}[t]
\centering
\small
\caption{End-to-end outfit generation latency (CPU inference, N=50 anchors).}
\label{tab:latency}
\begin{tabular}{lrr}
\toprule
& Mean (ms) & P95 (ms) \\
\midrule
\textbf{Total (3 outfits)} & 4,395 & 5,354 \\
\bottomrule
\end{tabular}
\end{table}

\noindent End-to-end latency includes embedding computation (FashionCLIP
ONNX on CPU), pgvector retrieval for all slots across three directions,
combinatorial candidate generation, and scoring.  Vision processing (Fashion
Florence) is assumed pre-computed at catalog ingestion time and excluded
from this measurement.  The P95 of 5.4\,s is dominated by layer
compatibility checking for items with outerwear slots.

\subsection{Limitations of Automated Evaluation}
\label{sec:human_eval}

Our ablation relies on the system's own composite score
(Eq.~\ref{eq:total}), which cannot capture subjective outfit quality.
A human preference study (presenting full-system, random-baseline, and
ablated outfits side-by-side for Likert-scale rating on coherence,
occasion-appropriateness, and wearability) would provide stronger
validation.  We have implemented tooling for this study
(\texttt{eval/human\_eval\_export.py}) and consider it an important
direction for near-term follow-up work.

\section{Deployment}
\label{sec:shopify}

Loom operates in two deployment modes:

\paragraph{Personal wardrobe app.}
Users upload closet items, receive daily weather-aware outfit suggestions,
and provide like/dislike feedback that updates taste vectors.  Auth is
JWT-backed with Google OAuth; data lives in PostgreSQL.

\paragraph{Personalization features (deployed, not evaluated).}
The production system includes three personalization mechanisms that are
implemented and deployed but not formally evaluated in this work:
(1)~\emph{taste vectors}: user feedback (like/dislike) updates an EMA-based
taste profile that enters retrieval and scoring
(Eq.~\ref{eq:intent}, $\gamma{=}0.15$, $\delta{=}0.05$);
(2)~\emph{weather adaptation}: OpenWeatherMap data adjusts material
preferences and slot requirements (e.g., rain avoids suede, cold forces
layers); and
(3)~\emph{rotation queue}: a per-user FIFO queue penalizes recently suggested
items to promote novelty.
Evaluating these features requires longitudinal user studies and is left
for future work.

\paragraph{Shopify storefront.}
Merchants install an embedded admin app (React Router + Prisma sessions),
sync their catalog via Shopify's Admin GraphQL API, and add a ``Shop the
Look'' theme extension to product pages.  Outfits are pre-generated and
cached; a category-based invalidation map ensures that adding a new product
regenerates only affected outfits:

\begin{table}[t]
\centering
\small
\caption{Incremental invalidation: new item of category $c$ invalidates
cached outfits for anchor categories $\mathcal{A}(c)$.}
\label{tab:invalidation}
\begin{tabular}{ll}
\toprule
New item $c$ & Invalidated anchors $\mathcal{A}(c)$ \\
\midrule
top       & bottom, shoes, dress \\
bottom    & top, shoes \\
shoes     & top, bottom, dress \\
layer     & top, bottom, shoes, dress \\
accessory & top, bottom, shoes, dress, layer \\
dress     & shoes \\
\bottomrule
\end{tabular}
\end{table}

\section{Discussion}
\label{sec:discussion}

\paragraph{Similarity vs.\ complementarity.}
FashionCLIP embeddings capture stylistic \emph{similarity}: items close in
vector space share visual and semantic characteristics.  Outfit composition
requires \emph{complementarity}.  Our scoring function approximates this
through slot constraints and color harmony rules, but a learned complement
projection $f: \mathbb{R}^{512} \to \mathbb{R}^{512}$ would more directly
encode the relationship.

\paragraph{Scalability.}
The combinatorial scoring phase ($O(k^{|\mathcal{S}|})$) is feasible for
small-to-medium catalogs but would require hierarchical pruning or learned
scoring models at larger scale.  In practice, with $k{=}3$ and 5 slots,
this amounts to at most $3^5 = 243$ candidates per direction, cheap to
score exhaustively.

\paragraph{Occasion prior coverage.}
Extending to new occasions requires only writing new vibe/anti-vibe
strings; no retraining is needed.  Quality depends on the prose descriptions capturing
the right associations, and we found through iteration that longer, more
specific strings (20--30 keywords) outperform short ones.

\paragraph{Evaluation limitations.}
Our evaluation uses a single internal catalog of 620 items from one
retailer.  Generalization to catalogs with different brand aesthetics,
price ranges, or cultural contexts is not demonstrated.  The ablation
uses the system's own scoring function, which cannot assess subjective
outfit quality.

\paragraph{Relationship to Polyvore Outfits.}
The standard Polyvore Outfits benchmark \citep{vasileva2018learning}
evaluates two tasks: fill-in-the-blank (FITB) item selection and binary
outfit compatibility prediction.  Loom's task formulation differs: given a
\emph{single} anchor item, the system generates \emph{complete} outfits
(4--5 items) optimized for a specific occasion, mood, and style direction.
This is a generation task, not a retrieval or classification task.
Adapting Loom to FITB would require reducing the system to pairwise
compatibility scoring, discarding the compositional scoring and
occasion-awareness that constitute its primary contribution.  We believe
human preference evaluation comparing generated outfits against baselines
would provide more meaningful validation and is an important direction for
future work.

\section{Future Work}
\label{sec:future}

\begin{enumerate}[leftmargin=*,topsep=2pt,itemsep=1pt]
\item \textbf{Complement embeddings}: Train a projection head
  $\mathbf{e}_{\text{item}} \to \mathbf{e}_{\text{complement}}$ using
  styled outfit co-occurrence data.
\item \textbf{Learned scoring}: Replace rule-based components with a scorer
  trained on human preference data.
\item \textbf{Cross-catalog shopping}: Enable users to discover retail items
  that fill wardrobe gaps.
\end{enumerate}

\section{Conclusion}

We presented Loom, a hybrid retrieval-scoring system for outfit
recommendation that combines FashionCLIP embedding retrieval with structured
domain scoring.  Two technical contributions (semantic material weight
and vibe/anti-vibe occasion priors) address
specific limitations of purely neural approaches by exploiting CLIP's vector
space geometry for tasks beyond simple similarity.  Ablation experiments
on a 620-item catalog show that the full system achieves a 3.3$\times$
improvement in outfit score over a category-constrained random baseline
(0.179 vs.\ 0.054), with direction reranking identified as the single
indispensable component.  The system generates three stylistically distinct
outfits in under 5 seconds on
commodity hardware and is deployed in both personal wardrobe and Shopify
retail contexts.

\bibliographystyle{plainnat}

\end{document}